\documentclass[AMA,STIX2COL]{MRM}
\articletype{Research Article}
\usepackage{useful}
\usepackage{graphicx}
\usepackage{placeins}
\usepackage{amssymb}
\usepackage[acronym,xindy,nopostdot,toc,section=chapter,nomain]{glossaries}

\newacronym{mri}{MRI}{Magnetic Resonance Imaging}
\newacronym{ct}{CT}{Computed Tomography}
\newacronym{pet}{PET}{Positron Emission Tomography}
\newacronym{pi}{PI}{Parallel Imaging}
\newacronym{cs}{CS}{Compressed Sensing}
\newacronym{rf}{RF}{Radio Frequency}
\newacronym{dl}{DL}{Deep Learning}
\newacronym{api}{API}{Application Programming Interface}
\newacronym{acs}{ACS}{Autocalibration Signal}
\newacronym{fmm}{FMM}{Fast Multipole Methods}
\newacronym{ift}{IFT}{Implicit Function Theorem}
\newacronym{dsm}{DSM}{Denoising Score Matching}
\newacronym[plural=DEQs,firstplural=Deep Equilibrium Network~(DEQs)]{deq}{DEQ}{Deep Equilibrium Network}
\newacronym[plural=MDEQs,firstplural=Multiscale Deep Equilibrium Network~(DEQs)]{mdeq}{MDEQ}{Multiscale Deep Equilibrium Network}
\newacronym{shine}{SHINE}{\textbf{SH}aring the \textbf{IN}verse \textbf{E}stimate}
\newacronym{fmri}{fMRI}{Functional MRI}
\newacronym[plural=T,firstplural=Teslas~(T)]{t}{T}{Tesla}
\newacronym{nmr}{NMR}{Nuclear Magnetic Resonance}
\newacronym{fid}{FID}{Free Induction Decay}
\newacronym{ft}{FT}{Fourier Transform}
\newacronym{ifft}{IFFT}{Inverse Fast Fourier Transform}
\newacronym{fft}{FFT}{Fast Fourier Transform}
\newacronym{dft}{DFT}{Discrete Fourier Transform}
\newacronym{ndft}{NDFT}{Nonuniform Discrete Fourier Transform}
\newacronym{nufft}{NUFFT}{Nonuniform Fast Fourier Transform}
\newacronym{psnr}{PSNR}{Peak Signal-to-Noise Ratio}
\newacronym{ssim}{SSIM}{Structural Similarity Index Measure}
\newacronym{mssim}{MSSIM}{Multiscale-SSIM}
\newacronym{snr}{SNR}{Signal-to-Noise Ratio}
\newacronym{knc}{KNC}{Koyasu Neurosurgical Clinic}
\newacronym{flair}{FLAIR}{Fluid Attenuated Inversion Recovery}
\newacronym{gre}{GRE}{Gradient Recalled Echo}
\newacronym{dwi}{DWI}{Diffusion-Weighted Imaging}
\newacronym{dct}{DCT}{Discrete Cosine Transform}
\newacronym{vds}{VDS}{Variable Density Sampling}
\newacronym{pdhg}{PDHG}{Primal-Dual Hybrid Gradient}
\newacronym{admm}{ADMM}{Alternating Direction Method of Multipliers}
\newacronym{ista}{ISTA}{Iterative Soft Thresholding Algorithm}
\newacronym{fista}{FISTA}{Faster ISTA}
\newacronym{pogm}{POGM'}{Proximal Optimal Gradient Method}
\newacronym{swi}{SWI}{Susceptibility Weighted Imaging}
\newacronym[plural=GPUs,firstplural=Graphical Processing Units~(GPUs)]{gpu}{GPU}{Graphical Processing Unit}
\newacronym[plural=GANs,firstplural=Generative Adversarial Networks~(GANs)]{gan}{GAN}{Generative Adversarial Network}
\newacronym{sgd}{SGD}{Stochastic Gradient Descent}
\newacronym{mlp}{MLP}{Multi-Layer Perceptron}
\newacronym{relu}{ReLU}{Rectified Linear Unit}
\newacronym{lrelu}{LReLU}{Leaky ReLU}
\newacronym{prelu}{PReLU}{Parametric ReLU}
\newacronym{gelu}{GELU}{Gaussian Error Linear Unit}
\newacronym[plural=CNNs,firstplural=Convolutional Neural Networks~(CNNs)]{cnn}{CNN}{Convolutional Neural Network}
\newacronym{pp}{P\&P}{Plug-and-Play}
\newacronym{dip}{DIP}{Deep Image Prior}
\newacronym{adni}{ADNI}{Alzheimer’s Disease Neuroimaging Initiative}
\newacronym{tse}{TSE}{Turbin Spin Echo}
\newacronym{oasis}{OASIS}{Open Access Series of Imaging Studies}
\newacronym{modl}{MoDL}{Model-based Deep Learning}
\newacronym{cg}{CG}{Conjugate Gradient}
\newacronym{gd}{GD}{Gradient Descent}
\newacronym{foe}{FoE}{Field of Experts}
\newacronym{rbf}{RBF}{Radial Basis Function}
\newacronym{nyu}{NYU}{New York University}
\newacronym{mwcnn}{MWCNN}{Multiscale Wavelet CNN}
\newacronym{radam}{RAdam}{Rectified ADAM}
\newacronym{dcp}{DCp}{Density Compensation}
\newacronym{nmse}{NMSE}{Normalized Mean Squared Error}
\newacronym{rmse}{RMSE}{Root Mean Squared Error}
\newacronym{mse}{MSE}{Mean Squared Error}
\newacronym{awgn}{AWGN}{Additive White Gaussian Noise}
\newacronym[plural=AFs,firstplural=Acceleration Factors~(AFs)]{af}{AF}{Acceleration Factor}
\newacronym{rss}{RSS}{Root-sum-of-squares}
\newacronym[plural=VAEs,firstplural=Variational Auto-Encoders~(VAEs)]{vae}{VAE}{Variational Auto-Encoder}
\newacronym{vi}{VI}{Variational Inference}
\newacronym{hmc}{HMC}{Hamiltonian Monte-Carlo}
\newacronym{uq}{UQ}{Uncertainty Quantification}
\newacronym[plural=DAEs,firstplural=Denoising Auto-Encoders~(DAEs)]{dae}{DAE}{Denoising Auto-Encoder}
\newacronym{cwgan}{cWGAN}{Conditional Wasserstein Generative Adversarial Network}
\newacronym{pdfs}{PDFS}{Proton-Density with Fat Suppression}
\newacronym{pd}{PD}{Primal-Dual}
\newacronym{eds}{EDS}{Expected Denoised Sample}
\newacronym{sde}{SDE}{Stochastic Differential Equations}
\newacronym{grappa}{GRAPPA}{Generalized Autocalibrating Partially Parallel Acquisitions}
\newacronym{qn}{qN}{Quasi-Newton}
\newacronym{lr}{LR}{Logistic Regression}
\newacronym{uli}{ULI}{Uniform Linear Independence}
\newacronym{opa}{OPA}{Outer-Problem Awareness}
\newacronym{nlp}{NLP}{Natural Language Processing}
\newacronym{rim}{RIM}{Recurrent Inference Machines}
\newacronym{mr}{MR}{Magnetic Resonance}
\newacronym{sense}{SENSE}{SENSitivity Encoding}
\newacronym[plural=CPUs,firstplural=Central Processing Units~(CPUs)]{cpu}{CPU}{Central Processing Unit}
\newacronym{dicom}{DICOM}{Digital Imaging and Communications in Medicine}
\newacronym{dc}{DC}{Data Consistency}
\newacronym{pcn}{PCN}{Parallel Coil Network}
\newacronym{storm}{SToRM}{SmooThness Regularization on Manifolds}
\newacronym{didn}{DIDN}{Deep Iterative Down-Up CNN}

\newacronym{spark}{SPARKLING}{Spreading Projection Algorithm for Rapid K-space sampLING}
\newacronym{more}{MORE}{Minimized Off Resonance Effect}
\newacronym{golf}{GoLF}{Gridding of Low Frequencies}
\newacronym{gs}{GS}{GoLF-SPARKLING}
\newacronym{tsd}{TSD}{Target Sampling Density}
\newacronym{adc}{ADC}{Analog to Digital Converter}
\newacronym{ute}{UTE}{Ultrashort Echo Time}
\newacronym{mrsi}{MRSI}{Magnetic Resonance Spectroscopy Imaging}
\newacronym{te}{TE}{Echo Time}
\newacronym{ti}{TI}{Inversion Time}
\newacronym{svd}{SVD}{Singular Value Decomposition}
\newacronym{uf}{UF}{Undersampling Factor}
\newacronym{tr}{TR}{Repetition Time}
\newacronym{fwhm}{FWHM}{Full Width at Half Maximum}
\newacronym{pnp}{PnP}{Plug-and-Play}
\newacronym{epi}{EPI}{Echo Planar Imaging}
\newacronym{psl}{PSL}{Peak-to-Sidelobe Level}
\newacronym{pnl}{PNL}{Peak-to-Noise Level}
\newacronym{nc}{NC}{Non-Cartesian}
\newacronym{psf}{PSF}{Point Spread Function}
\newacronym{tpsf}{TPSF}{Transform Point Spread Function}
\newacronym{loupe}{LOUPE}{Learning-based Optimization of the Under-sampling PattErn}
\newacronym{pns}{PNS}{Peripheral Nerve Stimulation}
\newacronym{tpi}{TPI}{Twisted Projection Imaging}
\newacronym{roi}{ROI}{Region of Interest}
\newacronym{mtf}{MTF}{Modulation Transfer Function}
\newacronym{floret}{FLORET}{Fermat Looped ORthogonal Encoded Trajectories}
\newacronym{twirl}{TWIRL}{TWisting Radial Lines}
\newacronym{fov}{FOV}{field-of-view}
\newacronym{caipi}{CAIPIRINHA}{Controlled Aliasing in Parallel Imaging Results in Higher Acceleration}
\newacronym{bjork}{BJORK}{B-spline parameterized Joint Optimization of Reconstruction and K-space trajectories}
\newacronym{pilot}{PILOT}{Physics-informed learned optimal trajectories}
\newacronym{projector}{PROJeCTOR}{PROjection for Jointly lEarning non-Cartesian Trajectories while Optimizing Reconstructor}
\newacronym[type=acronym]{cok}{CoK}{Center of K-space}
\newacronym[type=acronym]{kkt}{KKT}{Karush-Kuhn-Tucker}
\newacronym{tv}{TV}{Total Variation}
\newacronym[type=acronym]{mprage}{MP-RAGE}{Magnetization Prepared - RApid Gradient Echo}
\newacronym{tobs}{$T_{\rm Obs}$}{Observation time}

\newacronym{cp}{CP}{Cartesian Phyllotaxis}
\newacronym{pds}{PDS}{Poisson Disk Sampling}

\makenoidxglossaries

\newcommand{\spark}{\gls{spark}\ }

\renewcommand{\b}[1]{\mathbf{#1}}

\newcommand{\gsp}{\gls{golf}-\gls{spark}\ }

\newcommand{\capt}[2][]{\caption[{#1}]{ \textbf{{#1}}#2}}

\usepackage{gensymb}
\received{XXXX}
\revised{XXXX}
\accepted{XXXX}
\graphicspath{figures}
\begin{document}

\title{Combining Cartesian and non-Cartesian acceleration techniques with SPARKLING for 1mm isotropic whole-brain MPRAGE in a minute}

\author[1,2]{Chaithya G R}{\orcid{0000-0001-9859-6006}}
\author[3]{Aur\'elien Massire}{\orcid{0000-0002-4621-8254}}
\author[1,4,5]{Blanche Bapst}{\orcid{0000-0003-4958-7458}}
\author[1]{Alexandre Vignaud}{\orcid{0000-0001-9203-0247}}
\author[1,2]{Philippe Ciuciu}{\orcid{0000-0001-5374-962X}}

\authormark{Chaithya \textsc{et al}}

\address[1]{CEA, NeuroSpin, CNRS, Universit\'e Paris-Saclay, Gif-sur-Yvette, 91191, \country{France}}
\address[2]{Inria, MIND, Palaiseau, 91120, \country{France}}
\address[3]{Siemens Healthcare SAS, Courbevoie, 92400, \country{France}}
\address[4]{Department of Neuroradiology, AP-HP, Henri Mondor University Hospital, Créteil, 94000, \country{France}}
\address[5]{EA 4391, Université Paris Est Créteil, Créteil, 94000, \country{France}}

\corres{Chaithya G R PhD, NeuroSpin, CEA, Gif-sur-Yvette, \country{France} \email{chaithya.gr@cea.fr}}


\finfo{This work was granted access to the \fundingAgency{CCRT} High-Performance Computing (HPC) facility at CEA under the Grant \fundingNumber{CCRT2025-gilirach}. This work was granted access to the HPC resources of IDRIS under the allocation \fundingNumber{2021-AD011011153R4} made by \fundingAgency{GENCI}.}

\abstract[Summary]
{
    \section{Purpose}
$T_1$-weighted MPRAGE remains a cornerstone of clinical anatomical imaging, yet its long acquisition times constrain routine use. Established acceleration techniques, namely \gls{pi} and \gls{cs}, tend to introduce substantial noise and blurring when pushed to high acceleration factors. Although they rely on fundamentally different redundancies, combining them synergistically remains an open challenge.

    \section{Methods}
    The GoLF-SPARKLING framework was extended to jointly exploit two acceleration mechanisms: GRAPPA-based PI in the central k-space region and variable-density CS in the periphery, with independent acceleration factors in each zone. To preserve smooth signal evolution throughout the inversion-recovery period and avoid modulation artifacts, the acquisition trajectory was reordered accordingly. 
    The resulting method was evaluated prospectively in vivo at 1mm isotropic resolution and benchmarked against Wave-CAIPI and Poisson-disk sampling.

    \section{Results}
    The proposed hybrid approach produced sharper, less noisy, and more stable whole-brain images in approximately one minute than either acceleration strategy alone. 
    Purely PI-based reconstructions were degraded by high g-factor noise, while purely CS-based reconstructions exhibited pronounced blurring.
    Furthermore, this method yielded lower average volumetric errors in downstream automated brain segmentation than state-of-the-art acceleration techniques, demonstrating its clinical utility.

    \section{Conclusion}
    By jointly leveraging PI and CS, GoLF-SPARKLING achieves high acceleration factors that enable sub-minute, high-quality anatomical MRI. This translates into greater clinical throughput and more reliable imaging in patients who are challenging to scan.
}

\keywords{Parallel Imaging, Compressed Sensing, Ultrafast. SPARKLING}

\maketitle

\jnlcitation{\cname{%
\author{Chaithya G. R.}, 
\author{Aur\'elien Massire}, 
\author{Blanche Bapst},
\author{Alexandre Vignaud}, 
and 
\author{Philippe Ciuciu}}
(\cyear{2026}), 
\ctitle{Combining Cartesian and non-Cartesian acceleration techniques with SPARKLING for 1mm isotropic whole-brain MPRAGE in a minute}, \cjournal{Magn. Reson. Med.}, \cvol{2026;00:1--6}.}

\glsunset{grappa}
\glsunset{caipi}
\glsunset{gre}
\glsunset{te}

\glsreset{pi}
\glsreset{cs}
\section{Introduction}
$T_1$-weighted anatomical \gls{mri} has increasingly become a standard imaging modality in clinical practice, particularly for brain imaging as it provides high-resolution images of the brain's anatomy.
These images are essential for diagnosing various neurological conditions, planning surgeries, and monitoring disease progression \cite{Tae2025}.
Beyond routine scans, $T_1$-weighted sequences are increasingly integrated into emergency protocols, where the rapid assessment of acute neurological symptoms is critical \cite{Lang2024, Kazmierczak2020, Kendel2026}.
Although many emergency protocols currently rely on fast \gls{gre} based $T_1$ acquisitions, these often suffer from lower contrast compared to gold-standard sequences like \gls{mprage} \cite{mprage,Mugler1991} and its variants such as MP2RAGE \cite{mp2rage}.
Furthermore, high-resolution $T_1$-weighted imaging is vital for quantitative brain morphometry, which serves as a biomarker for neurodegenerative diseases \cite{Marek2025}.
In functional neuroimaging, \gls{mprage} provides the anatomical framework necessary for accurate tissue segmentation and precise localization of the BOLD signal.
Despite their clinical importance, these sequences remain time-consuming, often requiring 5 to 10 minutes to acquire a single volume.
Such lengthy acquisition times are challenging for patients who struggle to remain still, potentially leading to motion artifacts and reduced diagnostic quality.

To accelerate these acquisitions, two main strategies have been developed: \gls{pi}\cite{Deshmane2012} and \gls{cs}\cite{Lustig2007}.
\gls{pi} techniques are clinically established that use multi-coil acquisitions to exploit spatial redundancy, enabling image reconstruction from uniformly undersampled k-space data. Methods such as \gls{grappa}\cite{grappa}, SENSE\cite{sense}, and \gls{caipi}\cite{caipi} typically operate on Cartesian sampling patterns.
However, the acceleration achievable with \gls{pi} is fundamentally limited by the coil geometry and the signal-to-noise ratio penalty.
In many clinical settings, the use of hardware with a limited number of channels, such as 20-channel head coils, further restricts the practical acceleration factor.
In clinical context, even when higher-density arrays exist, 20-channel coils are often preferred because their more spacious design provides increased comfort and reduces patient anxiety.
Consequently, clinical use of \gls{pi} is usually limited to 2-4x speedups to avoid significant noise amplification and aliasing artifacts.

Extensions such as Wave-CAIPI\cite{mpragewavecaipi,wavecaipi}, which apply sinusoidal phase-encoding gradients during readout to evenly distribute aliasing across spatial directions, enable higher accelerations with negligible g-factor penalties.
In contrast, \gls{cs}\cite{Lustig2007} leverages image sparsity in a transform domain (e.g., wavelets) to reconstruct an MR image from incoherently undersampled k-space data. 
Since \gls{pi} and \gls{cs} exploit different forms of redundancy: spatial domain and transform-domain, respectively, they can be combined to achieve significantly higher acceleration than either method alone.

\begin{figure*}[!h]
    \centering
    \includegraphics[width=\textwidth]{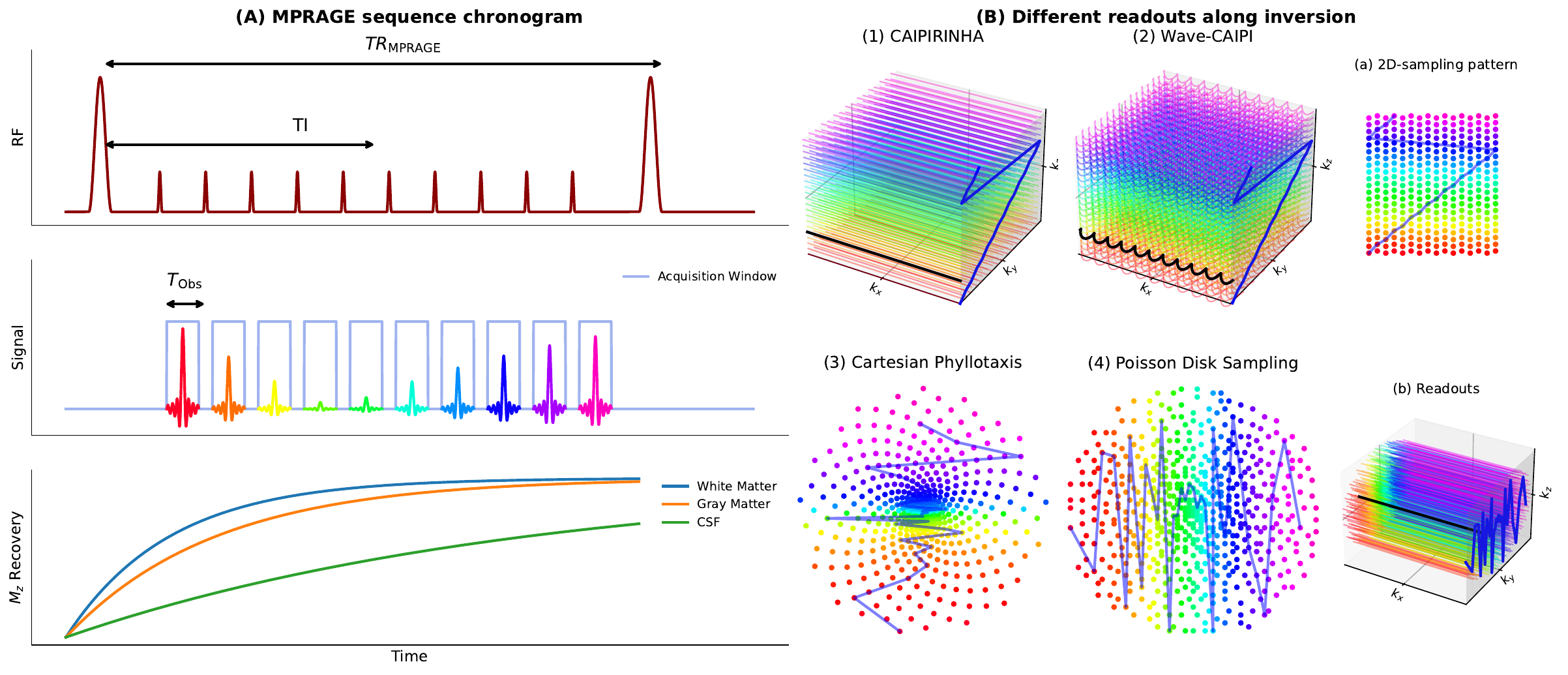}
    \capt[
        MPRAGE Sequence and different readouts along inversion color coded by inversion time :]{
        \label{fig:inversion_time} 
        \textbf{(A)} Sample chronogram of an \gls{mprage}\cite{mprage} sequence~ (top), where MR signal acquisition (middle) occurs during inversion recovery (bottom) following an inversion pulse. 
        The acquired k-space region depends on the applied gradients during readout as shown in \textbf{(B)}. 
        We show \textbf{(1)} CAIPIRINHA\cite{caipi} and its extension \textbf{(2)} Wave-CAIPI\cite{wavecaipi} sampling pattern with corkscrew readouts.
        2D sampling patterns for \textbf{(3)} Cartesian Phyllotaxis\cite{phyllotaxis} and \textbf{(4)} Poisson Disk Sampling along with \textbf{(b)} corresponding straight-line readout sampling patterns.
        For all the trajectories (1-3), we showcase trajectories traced in one inversion block in blue. 
        All acquisitions are color-coded by inversion time, from red (start of recovery) to purple (end) and one trajectory is highlighted in black.
    }
\end{figure*}

According to \gls{cs} theories\cite{puy2011variable,Chauffert_SIAM2014,adcock2017breaking,boyer2017compressed}, greater acceleration can be achieved via \gls{vds}, which can be efficiently achieved through the non-Cartesian imaging paradigm where acquisition occurs on flexible curves called trajectories rather than on straight lines.
Despite their scan time advantages, non-Cartesian in-out acquisition schemes are mainly limited to simple \gls{gre} sequences, due to repeated sampling of the k-space center. This results in acquired k-space data being sensitive to signal modulation in sequences such as \gls{mprage}, where magnetization evolves during acquisition, which can lead to inconsistent k-space weighting and image artifacts (Fig.\ref{fig:inversion_time}A).
Consequently, the application of \gls{cs}-based acceleration in anatomical \gls{mri} has largely been limited to sampling patterns with straight Cartesian readouts orthogonal to a 2D sub-sampling pattern like \gls{cp} or \gls{pds}\cite{phyllotaxis}.
However, such sampling patterns still remain suboptimal due to a lack of acceleration along the third dimension, namely the readout direction.
To mitigate these artifacts in the non-Cartesian imaging paradigm, MP-RAVE \cite{mprave} acquisitions constrained the sampling stack-of-stars pattern, with data during each inversion block still collected using Cartesian line-by-line readouts along a plane that rotates across inversions.
Extension to fully 3D radial acquisitions was carried out through MPnRAGE\cite{mpnrage} where reconstructions involved view-sharing of data across different \gls{ti} to obtain multiple inversion recovery images (Fig.\ref{fig:inversion_time}B-3).
Although these sampling schemes resulted in motion-robust scans\cite{mpnrage_motion}, the scan time remained relatively long due to repeated sampling of the k-space center beyond the Nyquist rate.

In this regard, the \spark algorithm was developed and later extended to \gls{golf} \cite{MOREGoLF} to optimize MR hardware compliant k-space sampling patterns for any specified \gls{vds}, with affine constraints to ensure cartesian sampling at the center of the k-space for efficient Nyquist rate sampling.
In this work, we utilize the \gsp trajectory framework and leverage hybrid acceleration: \gls{pi} with Cartesian acceleration at the center of the k-space and \gls{cs} through \gls{vds} sampling at high frequencies with non-Cartesian sampling for efficient coverage.
This novel technique will help bring commonly used multi-coil acceleration techniques like \gls{grappa} and \gls{caipi}\cite{grappa,caipi} to non-Cartesian imaging, which can be used independently to massively accelerate MR scans, enabling faster and higher resolution imaging, which will be beneficial for varied clinical applications. 
Although this method is generic and can be applied to any MR imaging modality, we present our results tailored to anatomical $T_1$-weighted acquisitions \gls{mri} with \gls{mprage}.
To achieve this, we re-order the acquisition trajectories during inversion recovery to ensure temporally smooth k-space sampling across the different \gls{ti}, to prevent the discussed signal modulation artifacts.

The remainder of the manuscript is organized as follows:
We begin in Section~\ref{sec:theory} with a short review of the anatomical sequence \gls{mri} and \gls{mprage}, followed by a brief overview of the developed \spark algorithm and crucially its extension \gls{golf}, which allows sampling the center of the k-space with Cartesian sampling.
Later, in Section~\ref{sec:methods} we introduce how such hybrid sampling of k-space opens doors for applying the non-Cartesian imaging paradigm to $T_1$-w \gls{mri} and how the use of \gls{grappa} and \gls{caipi} can help massively accelerate these acquisitions with little degradation in reconstructed image quality.
Finally, through prospective \textit{in vivo}  acquisitions, we proceed to test and benchmark our method through multiple ablation studies and compare our method against other state-of-the-art and clinically used acceleration techniques such as CAIPIRINHA\cite{caipi}, Wave-CAIPI\cite{wavecaipi}, \gls{cp} and \gls{pds}\cite{phyllotaxis}.
We show the potential benefit of using \gls{pi} and \gls{cs} in tandem to massively accelerate acquisitions, allowing for anatomical \gls{mri} with an isotropic resolution of 1mm for the whole brain.

\begin{figure*}[h]
    \includegraphics[width=\textwidth]{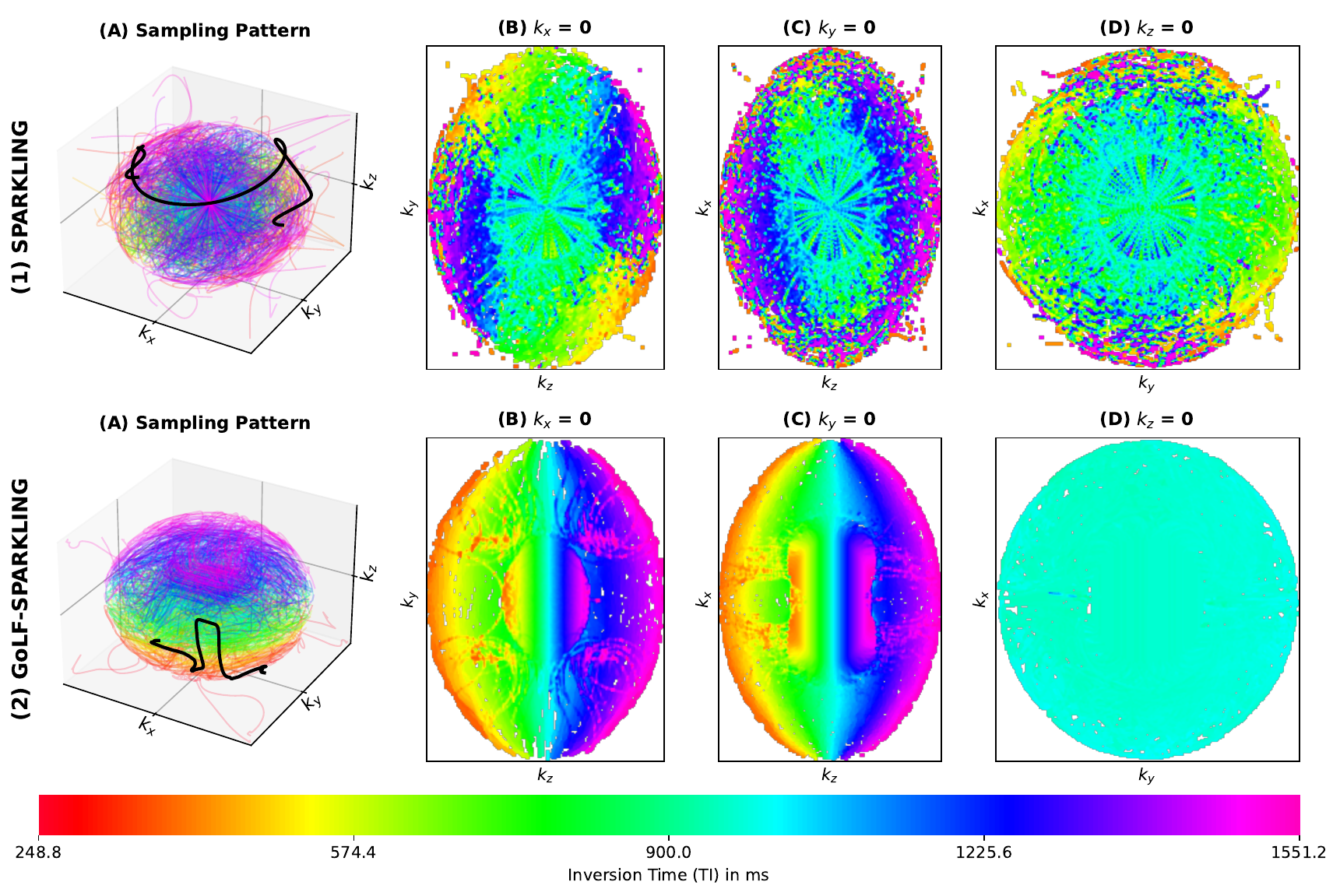}
    \capt[ Temporally smooth sampling of k-space across \gls{ti} achieved through \gsp ]{\label{fig:inversion_golf} \textbf{(A)} The 3D sampling patterns for \textbf{(1)} \spark and \textbf{(2)} \gsp trajectories, color coded based on the inversion time during which acquisition is performed, with central trajectory acquired at \gls{ti} = 900ms. We show one trajectory highlighted in black. We also show the mid-plane views of this trajectories along \textbf{(B)} $k_x = 0$, \textbf{(C)} $k_y = 0$ and \textbf{(D)} $k_z = 0$ planes.}
\end{figure*}

\section{Theory}
\label{sec:theory}
Following\cite{3dsparkling,MOREGoLF}, we consider 3D MRI with an isotropic field of view $\mathcal{F}$ and uniform resolution $N$, with the k-space domain normalized to $\Omega = [-1,1]^3$  by scaling spatial coordinates with $K_\textrm{max} = \tfrac{N}{2\mathcal{F}}$.
A conventional Cartesian scan requires $N^2$ segmented acquisitions with RF pulses spaced by \gls{tr}, giving a total acquisition time of $N^2 \times \textrm{TR}$, during which the data are collected over \gls{tobs}.
Accelerated scans use \gls{vds}, where the k-space is undersampled with \gls{tsd} $\rho(\b{x}):\Omega \rightarrow \mathbb{R}_+$ to densely sample the center of the k-space and reduce at higher frequencies.
In practice, the k-space is acquired with a realization of $\rho(\b{x})$, with reduced acquisition curves $N_c$, resulting in a \gls{af} given by $\frac{N^2}{N_c}$, which measures the speedup factor in actual MR scan time. 
Each trajectory is a curve in the k-space defined by $N_s = \lfloor\frac{T_\textrm{Obs}}{\Delta t}\rfloor$ discrete samples, $\b{k}[n] = (k^x[n], k^y[n], k^z[n])$, which is played at the gradient raster time $\Delta t$, while the acquired k-space $\mathbf{y}$ is sampled at \gls{adc} with dwell time $\delta t \leq \Delta t$ in $L$ coils, giving a data size of $L \times N_c \times N_s \frac{\Delta t}{\delta t}$.

\subsection{$T_1$-weighted anatomical \gls{mri} with \gls{mprage}}
$T_1$-weighted anatomical \gls{mri} acquisition protocols often involve a 180\textdegree\ RF pulse to flip the magnetization vector $M_z$, followed by imaging gradients applied as the magnetization recovers (Fig.\ref{fig:inversion_time}A).
To accelerate this process, rapid gradient echo readouts are typically grouped into trains with the number of readouts per train is defined as the turbo-factor (TF).
In particular, acquisition of \gls{mprage} involves $N_\textrm{tf}$ readouts per inversion segment, which are acquired centered around the inversion time \gls{ti} and this process is repeated every $\textrm{TR}_\textrm{MPRAGE}$.
The contrast in the images arises from the different rates at which tissues recover their magnetization, determined by their respective values of $T_1$ (Fig.~\ref{fig:inversion_time}A). 
Traditional Cartesian acquisitions involve $N_\textrm{tf} = N$ readouts along a $k_z$ plane acquired within one inversion block, with $N$ inversion blocks required for full coverage, resulting in a scan time of $N\times \textrm{TR}_\textrm{MPRAGE}$.
This ensures a smooth signal evolution in the k-space, resulting in reconstructed images without any artifacts associated with signal modulations during recovery.
MP-RAVE\cite{mprave} preserves this in-plane Cartesian acquisition while rotating the acquisition plane across inversions, yielding artifact-free reconstructions, thereby limiting the acquisition pattern to stack-of-stars.

Acceleration schemes inspired by \gls{pi} like \gls{caipi}\cite{caipi} (Fig.\ref{fig:inversion_time}B-(1)) and Wave-CAIPI\cite{mpragewavecaipi} (Fig.\ref{fig:inversion_time}B-(2)) have been proposed, which maintain such smooth signal evolutions, resulting in artifact free images.
However, acceleration inspired by \gls{cs} with \gls{vds} of the k-space has been limited to the application of a 2D undersampling pattern like \gls{pds} or \gls{cp}\cite{phyllotaxis} (Fig.\ref{fig:inversion_time}B-(3-4)) with Cartesian like straight line readouts (Fig.\ref{fig:inversion_time}B-(b)), to maintain temporal smoothness of the k-space signal with respect to inversion time.

\section{Methods}
\label{sec:methods}
\label{ssec:spark}
In this section, we present two major extensions to the original \spark framework that enable its application to inversion recovery imaging, specifically for $T_1$ weighted anatomical \gls{mri} using \gls{mprage} acquisitions.
We first outline the primary challenges in achieving this, followed by our proposed strategy to address them, and finally describe how we combine \gls{pi} and \gls{cs} to achieve massive acceleration.

To begin, we review that the original \spark algorithm \cite{Lazarus_MRM_19,3dsparkling} was designed to optimize a non-Cartesian k-space sampling pattern $\mathbf{K}=[\mathbf{k_i}]_{i=0}^{N_c}$ to match a prescribed \gls{tsd} $\rho(\mathbf{x})$.
Each individual trajectory $\mathbf{k_i} = [\mathbf{k_i}[n]]_{n=0}^{N_s}\in \Omega^{N_s}$ is constrained to be physically feasible and compatible with the gradient hardware of the scanner.
Overall, we optimize the 3D sampling pattern $\b{K}\in\Omega^{p}$ with $p=N_c\times N_s$ sample points using a weighted cost that combines an attraction term, which enforces adherence to \gls{tsd} $\rho(\b{x})$, and a repulsion term to prevent the clustering of samples\cite{3dsparkling,MOREGoLF}.

\subsection{Non-Cartesian anatomical \gls{mri} with \gsp}

Integration of non-Cartesian trajectories into inversion recovery sequences, such as \gls{mprage}, requires careful consideration of the temporal dimension.
In standard \spark, the optimization focuses on spatial distribution to match a target density, often neglecting the chronological order of acquisition. 
As illustrated in Fig.~\ref{fig:inversion_golf}-(1), this results in a discontinuous sampling of k-space relative to the Inversion Time \gls{ti}.
In panels (B) through (D) of the top row, the scattered color distribution reveals that the k-space center is sampled stochastically across the inversion block.
This temporal incoherence leads to significant signal modulation artifacts, as the high-energy components defining image contrast are captured at fluctuating points of the longitudinal magnetization recovery.

\begin{figure*}[h!]
    \includegraphics[width=\textwidth]{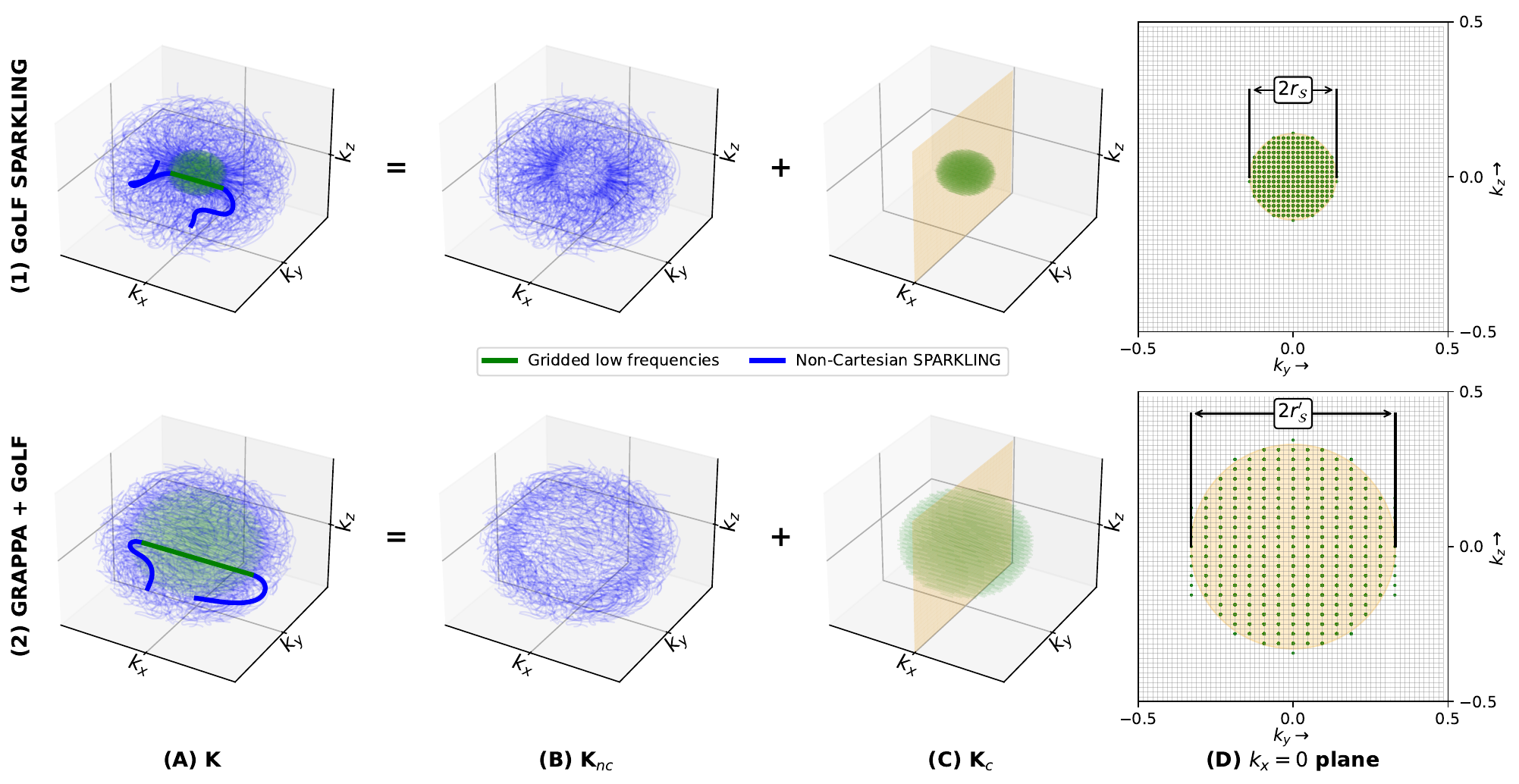}
    \capt[ \gsp and \gls{grappa} + \gls{golf} trajectories for $N_c=256$, with $N=64$ (for clearer visualization):]{\label{fig:golf_traj} \textbf{(1)} \gsp sampling pattern \textbf{(A)} $\b{K}$ consisting of \textbf{(B)} Non-Cartesian \spark at higher frequencies (blue) and \textbf{(C)} Gridded Cartesian sampling at k-space center (green). Corresponding mid-plane Cartesian sampling pattern along the $k_x = 0$ plane is shown in \textbf{(D)}. At the bottom row, the same split up of the sampling pattern is shown for  \gls{grappa} + \gls{golf}, where \gls{grappa} 2x2 acceleration is carried out in the k-space center, visible on the $k_x = 0$ slice view.}
\end{figure*}

To tackle this problem, we leverage \gls{golf}, an extension to the \spark framework originally proposed in \cite{MOREGoLF}.
Inspired by \gls{cs} theories that dictate dense sampling of the k-space center for superior image reconstruction, this approach enforces Cartesian sampling at low frequencies while maintaining non-Cartesian trajectories at higher frequencies to ensure excellent coverage with a variable density strategy.
The Cartesian sampling is achieved by applying trajectory specific affine constraints that guarantee optimal Nyquist rate sampling at the center of the spectrum.
Specifically, for a sampling pattern comprising $N_c$ trajectories, we enforce Cartesian grid alignment within a central sphere $\mathcal{S}$ of radius $r_\mathcal{S}$, as illustrated in Fig.~\ref{fig:golf_traj}-(1D), where from Sec 3.2 in\cite{MOREGoLF} we get:
\begin{align}
    r_{\mathcal{S}} = \frac{2}{N}\left\lfloor\sqrt{\frac{N_c}{\pi}}\right\rfloor
\end{align}

While this sampling strategy was primarily designed to ensure optimal Nyquist rate sampling at the k space center and dispatch remaining samples to high frequencies, it also provides a mechanism to smoothly distribute the acquisition across successive inversion blocks.
Instead of redundantly acquiring the entire center of k space during every single shot, successive inversion blocks collect interleaved subsets of the central sphere $\mathcal{S}$.
This interleaved approach ensures that the critical central k space data are sampled more uniformly across the inversion recovery curve.

The resulting temporal smoothness is clearly visible in Fig.~\ref{fig:inversion_golf}-(2), where the k-space center is acquired at a stable $TI \approx 900$ms.
By playing the trajectories in a sequence similar to Cartesian sampling at the center, the acquisition achieves the temporal smoothness necessary for high-fidelity anatomical MRI. 
As seen in the mid-plane views (B, C, and D) of the bottom row, the smooth color gradients indicate that the k-space is filled in a continuous manner relative to the TI. 
Specifically, the central trajectory is acquired at the null point or desired contrast point (e.g. TI = 900 ms), while the peripheral regions are filled progressively. 
This transition significantly reduces jumps in magnetization between neighboring k-space samples, allowing the use of \gls{vds} to accelerate acquisitions by a factor of $\text{AF} = N\lceil\frac{N_{\text{tf}}}{N_c}\rceil$ without sacrificing image contrast stability or clarity.

\renewcommand{\gsp}{\gls{gs}\ }
\subsection{Accelerating \gls{golf} with \gls{grappa}}

\gsp introduces a novel compound sampling approach to measure the k-space with trajectories having both Cartesian and non-Cartesian parts to extract the best of both worlds.
In this section, we present how this sampling pattern allows its application in inversion recovery imaging and to bring clinically established methods \gls{pi} such as \gls{grappa} \cite{grappa} and \gls{caipi}\cite{caipi} for additional acceleration.
Later, we discuss the general acquisition paradigm under which we shall proceed to carry out prospective \textit{in vivo} acquistions, and how the acquired k-space data are reconstructed.
Finally, we discuss how all the techniques for acquiring and reconstructing data in this work are available to all to promote reproducible research and open science.

\begin{figure}[h!]
    \includegraphics[width=0.5\textwidth]{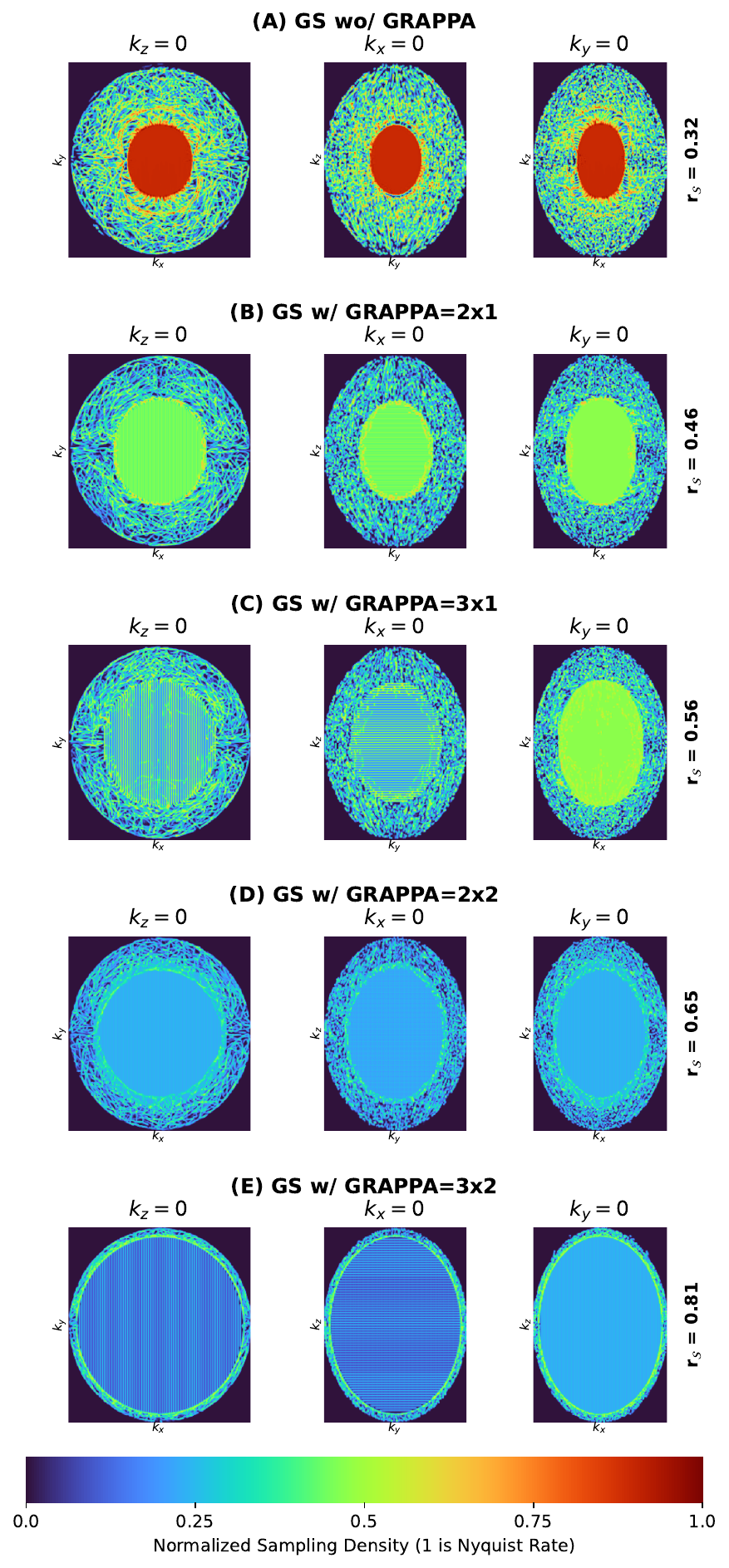}
    \caption{\label{fig:golf_with_grappa_traj} K-space sampling density for \textbf{(A)} vanilla \gsp (GS) trajectories without GRAPPA and with GRAPPA \textbf{(B)} $2\times 1$, \textbf{(C)} $3\times 1$, \textbf{(D)} $2\times 2$ and \textbf{(E)} $3\times 2$. We show the sampling densities along mid-planes $k_z= 0$ (left row), $k_x = 0$ (middle row) and $k_y = 0$ (right row). The corresponding color bar for the sampling density is shown at the bottom which is normalized by Nyquist sampling rate.}
\end{figure}

The center of space k in the \gsp trajectories is gridded at the full Nyquist sampling rate, which can potentially be accelerated through the use of popular clinically established \gls{pi} techniques such as \gls{grappa}/SENSE \cite{grappa,sense} and \gls{caipi}\cite{caipi} (Fig.\ref{fig:golf_traj}-(2)).
This enables acceleration via two complementary strategies: Uniform undersampling of the Cartesian-sampled k-space center and non-uniform undersampling at higher frequencies with \gls{vds} through the non-Cartesian imaging paradigm.
In this way, we exploit spatial redundancy in multi-coil data to reconstruct the center of the k-space, while relying on the sparsity of the transform-domain \gls{cs} to recover high-frequency details.
In practice, this is implemented by modifying the trajectory-specific affine constraints to uniformly skip samples at the k-space center, resulting in a subsampling pattern similar to \gls{grappa}.

To show the benefits of the proposed update, we generate \gsp trajectories with $N_c = 4624$ using $2 \times 1$, $3 \times 1$, $2 \times 2$, and $3 \times 2$ \gls{grappa} acceleration factors.
The corresponding sampling density profiles along the $k_x = 0$, $k_y = 0$, and $k_z = 0$ midplanes are illustrated in Fig.~\ref{fig:golf_with_grappa_traj}.
Keeping a constant scan time, the fraction of the k-space center covered by Cartesian sampling increases from $r_\mathcal{S} = 0.32$ without \gls{grappa} to $0.46$, $0.56$, $0.65$, and $0.81$ for the respective \gls{grappa} accelerated configurations.
In the limiting case, employing a $3 \times 3$ \gls{grappa} acceleration with $N_c = 4624$ results in a purely uniformly undersampled Cartesian acquisition.

This hybrid sampling framework provides a tunable mechanism to balance the strengths of \gls{pi} and \gls{cs} by adjusting the transition between the Cartesian and non-Cartesian regimes. 
Using uniform undersampling within the central k-space sphere $\mathcal{S}$, we leverage spatial redundancy and multi-coil sensitivity profiles for \gls{grappa}-based reconstruction, which is particularly robust for recovering high-SNR contrast information. 
Conversely, high-frequency regions are sampled using non-Cartesian \gsp trajectories, which maximize the efficiency of \gls{vds} and rely on transform-domain sparsity and \gls{cs} priors to recover fine structural details.
This dual-strategy allows for finding an optimal regime between these two reconstruction paradigms. 

We proceed to test the added value of our method through prospective \textit{in vivo} acquisitions, where we compare the quality of reconstruction across a range of sampling configurations.
Specifically, we evaluate the transition from a standard \gsp pattern without \gls{grappa} acceleration to configurations incorporating $2 \times 1$, $3 \times 1$, $2 \times 2$, and $3 \times 2$ \gls{grappa} factors at the center of k-space.
To provide a controlled comparison, all acquisitions are performed at the same global \gls{af}, effectively redistributing the ``acceleration budget'' between \gls{pi}-based Cartesian sampling at the center and \gls{cs}-based non-Cartesian sampling at the periphery.
This comparison allows us to identify the optimal balance between these two reconstruction paradigms for anatomical imaging.

\subsection{Benchmarking massively accelerated almost-a-minute \gls{mri}}
To evaluate the performance of the proposed \gsp framework in the regime of extreme acceleration, we conducted a comprehensive benchmark study against established clinical and research standards.
We compare our hybrid sampling approach against other state-of-the-art techniques which use purely \gls{pi}-based methods or purely \gls{cs}-based trajectories under highly constrained scan times.
For the primary benchmark, all methods were restricted to a 62-second acquisition window to represent the ``almost-a-minute'' imaging paradigm, where we scan a clinical routine involving a whole brain $T_1$-weighted \gls{mprage} at 1mm isotropic resolution.
In this category, \gsp was compared with \gls{caipi} $3 \times 3$ \cite{caipi} and Wave-CAIPI $3 \times 3$ \cite{wavecaipi}, representing the state-of-the-art in Cartesian \gls{pi}.
Additionally, we compared with purely \gls{cs}-driven strategies, specifically \gls{pds} and \gls{cp} sampling \cite{phyllotaxis}, which use incoherent aliasing for artifact-free image reconstruction.

Note that, due to lack of availability standard sequences for CAIPIRINHA and Wave-CAIPI, we proceeded to implement them on our own. 
For both of these methods, we ensured temporal sampling throughout inversion (see Fig.~\ref{fig:inversion_time}-(1-2)).
For Wave-CAIPI, we implemented\cite{mpragewavecaipi} with a wave amplitude of 8.8 mT/m and 11 sinusoidal cycles per readout.
For the \gls{cp} and \gls{pds} schemes, we used a Research Application sequence developed by the manufacturer~(Siemens-Healthineers).

To ensure a fair comparison with established Cartesian acceleration techniques, we further evaluated \gls{caipi} $3 \times 2$ and Wave-CAIPI $3 \times 2$ using elliptical sampling of the k-space.
Since these configurations resulted in a slightly longer scan time of 75 seconds (representing a 25\% increase in data compared to the 62-second acquisitions), we re-generated all other trajectories in our benchmark at the exact same 75-second mark.
This dual-baseline approach allows for a rigorous assessment of whether the performance gains of our method stem from the sampling trajectory itself rather than differences in the total amount of acquired data.

\subsection{Acquisition paradigm}
\label{sec:acq_params}
We carried out prospective \textit{in vivo} acquisitions on a clinical 3T MR system (Magnetom Cima.X, Siemens Healthineers, Forchheim, Germany) to validate all our results and compare them against other state-of-the-art approaches.
In all experiments, the k-space data were acquired in sagittal orientation with a target resolution of $1 \text{mm}^3$ isotropic, with an image matrix size of $256\times240\times176$, leading to total brain coverage, which is regularly used in clinical practice.
For our baseline, we used fully sampled Cartesian \gls{mprage} acquisition with $N_\textrm{tf} = 176$ and $\textrm{TR}_\textrm{MPRAGE}$ = 2.3 s, leading to a scan time of nearly 9 minutes.
The inner imaging readouts were acquired with a flip angle of $9\deg$ and \gls{te}/\gls{ti}/\gls{tr} was 3ms/900ms/8ms and BW = 240Hz/px (\gls{tobs}=4.16ms).
For all accelerated acquisitions, we matched the same parameters of the acquisition protocol to ensure the same contrast, with subsampling carried out with only $N_c$ trajectories, which was set based on the desired \gls{af} / scan time.
An external \gls{acs} scan was performed in which the central $24\times 24$ k-space was acquired at a minimum \gls{te} and \gls{tr} to estimate sensitivity maps, which is later used during reconstructions.
While, the k-space data was acquired with \gls{tobs}= 4.16ms (to match BW=240Hz/px), leading to $N_s=416$ and data was sampled by the \gls{adc} at $\delta t = 5$ \textmu s.
All scans were performed on two healthy volunteers with the approval of local and national ethical committees for the protocol (CPP 100050), and after the written consent was obtained.

\subsubsection{Ablation study with different coil configurations}
To perform an ablation study of massively accelerated acquisitions across different coil geometries, we conducted experiments using 64-channel and 20-channel manufacturer head/neck coils.
This comparison allows us to evaluate the robustness of different acceleration techniques in varying degrees of coil-based spatial encoding.
By testing in high-density and standard clinical arrays, we assess how the interplay between \gls{pi} and \gls{cs} scales with the available \gls{snr} and coil sensitivity profiles.

\begin{figure*}
    \centering
    \includegraphics[width=\textwidth]{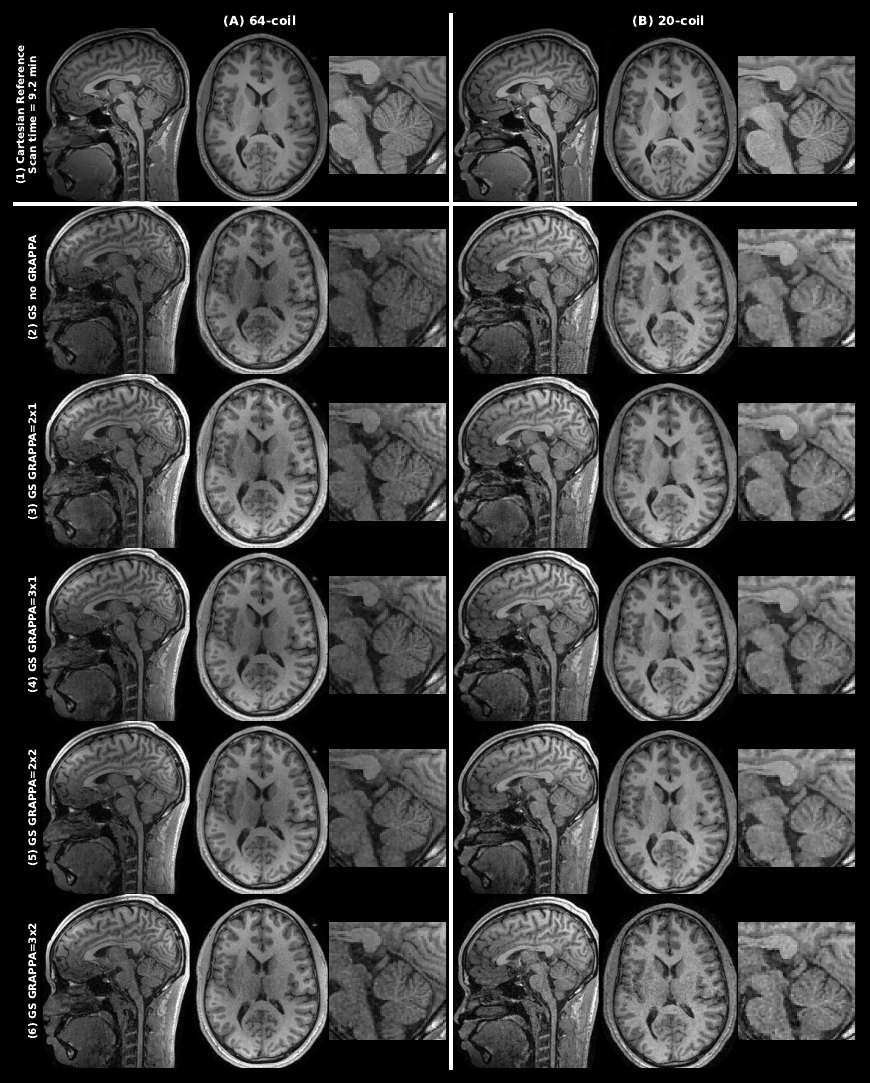}
    \caption{\label{fig:golf_vs_ggrappa} \gls{tv}-regularized reconstructions for \textbf{62-second} acquisitions comparing \textbf{(2)} \gsp without \gls{grappa} to \textbf{(3--6)} varying central \gls{grappa} acceleration in k-space center.
    A \textbf{(1)} 9.2-minute fully sampled Cartesian acquisition is provided as a reference.
    Acquisitions were performed using \textbf{(A)} 64-channel and \textbf{(B)} 20-channel coils to test robustness.
    Each panel displays sagittal (left) and axial (middle) views, with a zoomed cerebellum inset (right).
    The green, yellow, and red arrows indicate good, okay, and bad reconstructions, respectively.
}
\end{figure*}

\subsection{Image reconstruction}
\label{sec:recon}
We initiate reconstruction by estimating coil sensitivity maps from the \gls{acs} data using the ESPiRIT algorithm~\cite{uecker2014espirit}.
To minimize computational complexity in terms of both memory and processing speed, sensitivity maps were initially estimated at a four-fold downsampled resolution, as implemented in mri-nufft~\cite{mrinufft}\footnote{\url{https://mind-inria.github.io/mri-nufft/generated/autoexamples/extras/example_smaps.html}} package.
During the final reconstruction stage, these maps are upsampled back to the target image matrix size.
Since \gls{acs} is a brief 1-second acquisition performed at the start of the scan, this estimation can be executed during subsequent accelerated acquisition.
This parallelized workflow significantly reduces the overall reconstruction latency for the clinical user.

All acquired k-space data $\b{y}_\ell$ from different sampling patterns were reconstructed through iterative reconstruction with a Total Variation~(\gls{tv}) regularization:
\begin{equation}
    \b{\hat{x}} = \arg \min_{x\in \mathbb{C}^M} \frac{1}{2L} \sum_{\ell=1}^L|| \mathcal{F}_{\b{K}} S_\ell x  - \b{y}_\ell||_2^2 + \lambda ||\nabla \mathbf{x}||_{1,2}
\end{equation}
where $\nabla$ is the linear operator of finite differences~(over neighboring voxels) and the 2-norm is taken along the finite differences.
In practice, this reconstruction is performed using a primal-dual optimization scheme~\cite{Condat2013}.
The regularization strength $\lambda$ is manually fine tuned by grid-search for each of the undersampling patterns.

Image reconstruction was performed on a 32-core AMD Milan @ 2.45GHz CPU and an NVIDIA A100 GPU, achieving a reconstruction time of approximately 1.3 seconds per iteration.
The algorithm converges within 25 iterations or nearly 30 seconds when the relative change between consecutive iterates drops below $10^{-3}$, enabling direct reconstruction in the scanner.

We note that while a wide range of complex, deep learning based learned priors can enhance image quality and accelerate reconstructions, we have not applied them to any of our results to be able to efficiently judge the performance of the acquisitions, purely from an acquisition standpoint.

\subsection{Reproducible research and Open Science}
To ensure reproducibility, the k-space sampling trajectories and Gadgetron reconstruction workflows will be open sourced through MRI-NUFFT~\cite{mrinufft} and PySAP-MRI~\cite{pysap}packages.
For all our experiments, we played these trajectories on the Siemens scanner through our MR sequence \texttt{ns \_tfl \_arb}, where any arbitrary k-space trajectory can be played during readout, allowing for fast prototyping of the non-Cartesian imaging paradigm for different \gls{gre} and inversion recovery based imaging modality.
Additionally, our Gadgetron reconstructions are also translated to Siemens framework for image reconstruction environments (FIRE), allowing for direct use of our trajectories in clinical setup with live online reconstructions.
Both the MR pulse sequences that we developed and the corresponding FIRE reconstructions are shared through the Siemens C2P exchange platform.
Finally, the \spark~\cite{3dsparkling,MOREGoLF} codes remain available upon request to the broader research community for non-commercial research use to optimize trajectories tailored to other acquisition paradigms.

\begin{figure*}
    \centering
    \includegraphics[width=\textwidth]{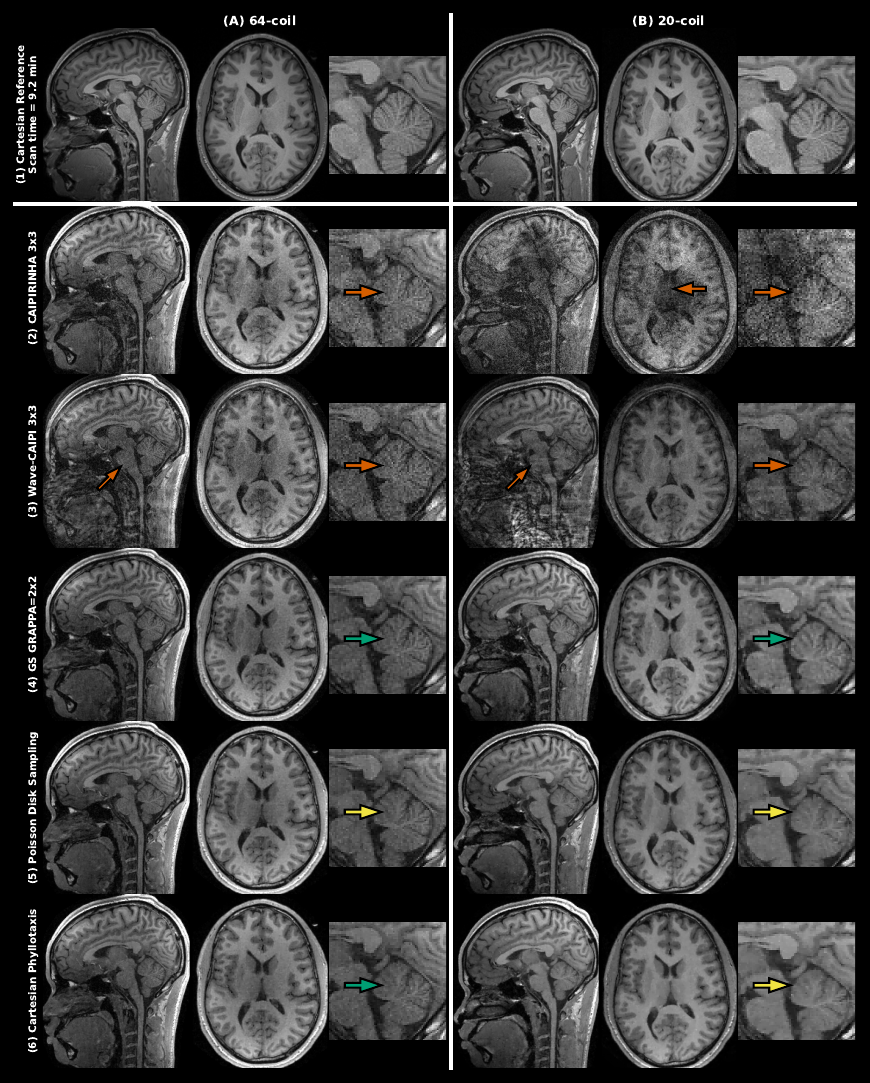}
    \caption{\label{fig:ggrappa_vs_sota} \gls{tv}-regularized reconstructions for \textbf{62 second} scan comparing \textbf{(4)} \gsp $2\times 2$ against state-of-the-art methods including \textbf{(2)} \gls{caipi} $3\times 3$, \textbf{(3)} Wave-CAIPI $3\times 3$, \textbf{(5)} Poisson Disk, and \textbf{(6)} Cartesian Phyllotaxis with \textbf{(1)} 9.2-minute fully sampled Cartesian reference.
    Acquisitions using \textbf{(A)} 64-channel and \textbf{(B)} 20-channel coils evaluate robustness across coil geometries.
    Each panel displays sagittal (left) and axial (middle) views with a zoomed cerebellum inset (right).
    The green, yellow, and red arrows indicate good, okay, and bad reconstructions, respectively.
}
\end{figure*}

\section{Results}
\label{sec:results}
We proceed to present the results for prospectively accelerated 1 minute anatomical \gls{mri}.

\subsection{Benefit of \gls{grappa} in \spark}

In Fig.~\ref{fig:golf_vs_ggrappa}, we show the impact of integrating Cartesian undersampling based on \gls{grappa} within the \gsp framework.
When there is no \gls{grappa}, the reconstructed images appear notably blurry, failing to capture fine anatomical details.
By increasing the \gls{grappa} factor at the center of k-space, a larger portion of the ``sampling budget'' is redirected toward higher spatial frequencies, with uniform undersampling of the center of k-space (see Fig.~\ref{fig:golf_with_grappa_traj}).
This shift allows the non-Cartesian \gls{vds} to cover a wider extent of the k-space, resulting in significantly sharper reconstructions.
This improvement in high-frequency detail is particularly visible in the zoomed insets of the cerebellum, where the folia become increasingly distinct as the amount of centeral \gls{grappa} acceleration increases.

However, higher \gls{pi} factors also introduce noise amplification in the center of the brain due to the g-factor.
This effect is clearly visible in the $3 \times 2$ \gls{grappa} acquisitions for the 64-coil and is even more pronounced for the 20-coil due to its limited spatial encoding capabilities.
We observed that the $2 \times 2$ \gls{grappa} accelerated acquisition embedded within GS provides the optimal tradeoff between \gls{pi} noise levels and \gls{cs} detail recovery at high frequencies.

\begin{figure*}
    \centering
    \includegraphics[width=\textwidth]{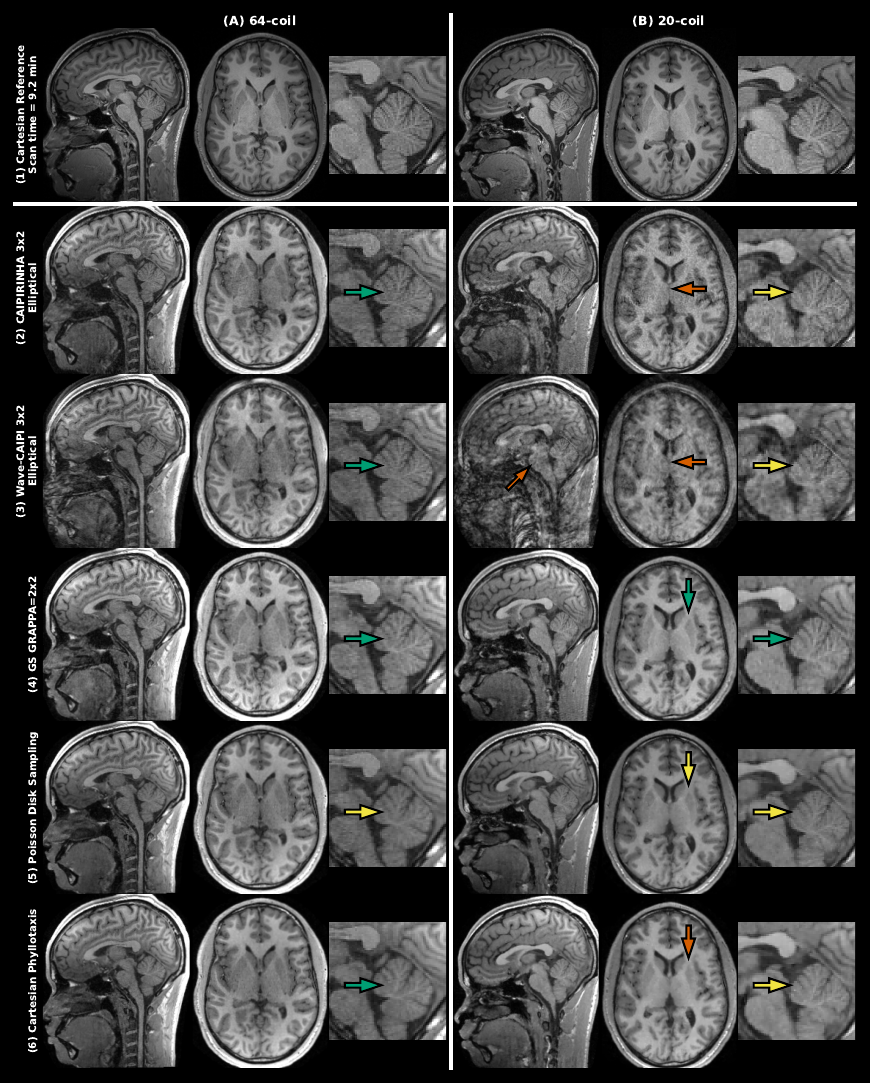}
    \caption{\label{fig:ggrappa_vs_sota_circ}  \gls{tv}-regularized reconstructions for \textbf{75 second} scan comparing \textbf{(4)} \gsp $2\times 2$ against state-of-the-art methods including \textbf{(2)} Elliptical \gls{caipi} $3\times 2$, \textbf{(3)} Elliptical Wave-CAIPI $3\times 2$, \textbf{(5)} Poisson Disk, and \textbf{(6)} Cartesian Phyllotaxis with \textbf{(1)} 9.2-minute fully sampled Cartesian reference.
    Acquisitions using \textbf{(A)} 64-channel and \textbf{(B)} 20-channel coils evaluate robustness across coil geometries.
    Each panel displays sagittal (left) and axial (middle) views with a zoomed cerebellum inset (right).
    The green, yellow, and red arrows indicate good, okay, and bad reconstructions, respectively.
    }
\end{figure*}

\begin{figure*}
    \centering
    \includegraphics[width=\textwidth]{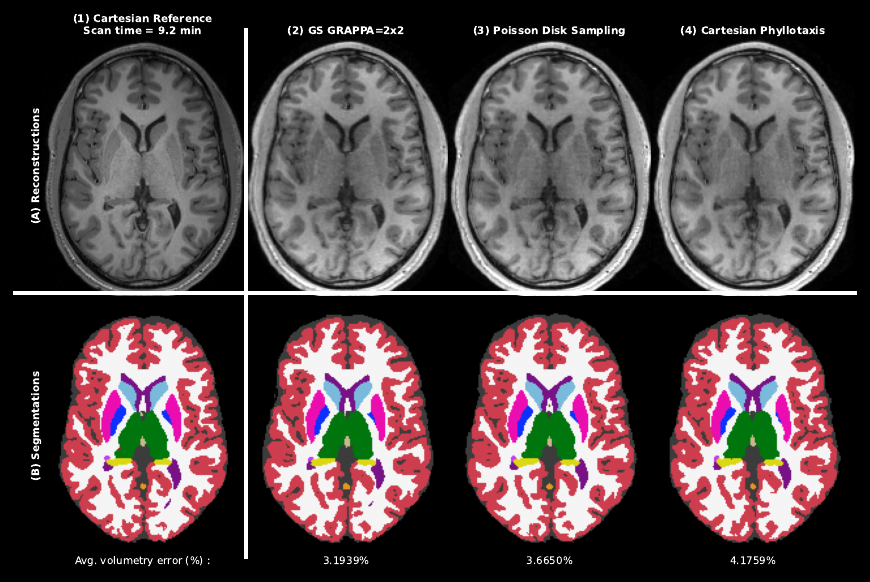}
    \caption{\label{fig:ggrappa_vs_sota_seg} \textbf{(A)} Axial views of \gls{tv}-regularized reconstructions for \textbf{62 second} scan acquired on 64-channel receiver. We compare \textbf{(2)} \gsp $2\times 2$ against state-of-the-art methods including \textbf{(3)} Poisson Disk, and \textbf{(4)} Cartesian Phyllotaxis with \textbf{(1)} 9.2-minute fully sampled Cartesian reference.
    \textbf{(B)} The corresponding segmentation masks from FreeSurfer SynthSeg.
    The corresponding average percentage error of volumetric tissue quantification are noted below.}
\end{figure*}

\subsection{Benchmarking massively accelerated anatomical \gls{mri}}
\label{ssec:benchmark}

In Fig.~\ref{fig:ggrappa_vs_sota}, we evaluate the performance of the proposed \gsp $2\times 2$ framework against established clinical and research benchmarks for a 62-second acquisition. 
To ensure a fair comparison with Cartesian methods, we also present results for a 75-second acquisition in Fig.~\ref{fig:ggrappa_vs_sota_circ}, where \gls{caipi} and Wave-CAIPI are performed using elliptical sampling and a $3 \times 2$ acceleration factor.

Reconstructions using only \gls{pi} acceleration, such as \gls{caipi} $3\times 3$, exhibit significantly noisy reconstructions.
Although such noise is mildly present in the 64-coil data, it becomes severe in the 20-coil acquisitions, which are typically favored in clinical practice due to the increased cost of the 64-coil configuration and the lack of comfort for patients with larger head sizes.
Although Wave-CAIPI reconstructions appear less noisy than standard \gls{caipi}, they remain heavily artifacted, particularly in the sagittal view.
These artifacts likely stem from a lack of auto-PSF calibration for wave gradients~\cite{mpragewavecaipi} on the scanner due to the unavailability of the corresponding proprietary sequences and codes. 
However, Wave-CAIPI still does not produce diagnostic quality in the 20-coil configuration.

Purely \gls{cs}-based techniques, including \gls{pds} and \gls{cp}, provide more stable and less noisy reconstructions compared to Cartesian \gls{pi}.
However, these methods suffer from noticeable blurring, as evidenced by the loss of fine structural detail in the zoomed cerebellum insets.
In contrast, the \gsp trajectories with $2\times 2$ \gls{grappa} acceleration provide an optimal tradeoff between the structural fidelity of \gls{pi} and the denoising capabilities of \gls{cs}.
This hybrid approach yields stable, less noisy, and sharper results in both the 64-coil and 20-coil configurations, demonstrating superior robustness to varying coil geometries.

Finally, we evaluate the performance of the proposed method in downstream tasks, as automated brain segmentation and volumetry are critical use cases for the T1 weighted sequence \gls{mprage} in clinical and research settings.
Maintaining high structural fidelity in T1 weighted MR images is essential to ensure that volumetric measurements remain faithfully representative of the underlying brain anatomy for longitudinal monitoring or diagnostic purposes.
We segmented the reconstructed T1 weighted MR images using FreeSurfer SynthSeg~\cite{synthseg,synthseg_freesurfer} and subsequently performed volumetry on the resulting segmentation masks.
Relative percentage errors are calculated with respect to the volumetric values of the Cartesian reference, with the average error reported.
Due to inter scan motion, direct DICE score metrics\cite{dice} are not suitable, as even post hoc alignment of segmented images would likely retain residual errors.
As illustrated in Fig.~\ref{fig:ggrappa_vs_sota_seg}B, the proposed \gsp $2\times 2$ framework achieves a lower average volumetric error compared to other state of the art acceleration techniques.

\section{Discussion and Conclusion}
\label{sec:discon}
The results of this study demonstrate that the \gsp framework provides a highly tunable mechanism to steer the integration of \gls{pi} and \gls{cs} in massively accelerated anatomical \gls{mri}.
By partitioning k-space into a Cartesian center and a non-Cartesian periphery, we effectively leverage multi-coil sensitivity profiles and transform-domain signal sparsity as complementary priors.
Our comparative analysis reveals a distinct tradeoff in which purely \gls{pi}-based methods often suffer from significant g-factor noise amplification, and purely \gls{cs}-based techniques tend to produce stable but overly blurry reconstructions.
The $2 \times 2$ \gls{grappa}-accelerated \gsp configuration represents an optimal regime in this tradeoff, yielding 3D MR images that are sharper than traditional \gls{cs} and more resistant to noise than standard Cartesian \gls{pi}.

During the design and testing of these novel hybrid trajectories, we did not perform an extensive analysis of trajectory errors associated with complex gradient profiles, as this investigation was previously detailed in the original 3D \spark study \cite{3dsparkling}.
Throughout our repeated evaluations, the Siemens systems demonstrated robustness against eddy current effects, with the measured trajectory errors remaining sufficiently low to result in any substantial artifacts in the reconstructed MR images.
In this work, we did not employ complex deep learning based priors for image reconstruction, which could further enhance image quality.
Instead, we restricted our study to well understood conventional priors such as \gls{tv} to ensure mathematical guaranties and avoid hallucination artifacts, facilitating a more direct clinical deployment.
Furthermore, employing deep learning reconstructions could obscure the performance gap between different sampling patterns, confounding a direct comparison of the trajectories themselves.
Future studies will integrate state-of-the-art deep learning reconstruction techniques, including plug-and-play and diffusion priors~\cite{kamilov2023plug,pnp_comby_mri,chung2023solving}.
Additionally, the proposed hybrid acceleration strategy will be incorporated into data-driven trajectory design frameworks such as PROJeCTOR~\cite{radhakrishna2023jointly}.
This will allow for the joint optimization of the sampling pattern and the reconstruction network on large scale datasets like the Calgary brain dataset~\cite{calgary}. 

Regarding the benchmarks, it is important to note that the performance of Wave-CAIPI might be further improved by using auto-calibrated \gls{psf} reconstructions, as proposed in \cite{mpragewavecaipi}.
The artifacts observed in our implementation likely stem from the lack of precise wave-gradient trajectory modeling, which is critical for maintaining coherence in Cartesian wave-readouts.
Nevertheless, even without such specialized calibration, our hybrid approach demonstrated superior robustness, particularly when using limited coil geometries such as the 20-channel array.

This \gsp approach can be naturally extended to MP2RAGE~\cite{mp2rage} and other multi-contrast sequences MP2RAGEME\cite{Caan2018} and multi-echo \gls{gre}.
In such applications, complementary sampling patterns could be leveraged in different \gls{ti} and \gls{te} to maximize the efficiency of the coverage of the k-space.
This would build on the concepts introduced in MPnRAGE~\cite{mpnrage}, but would significantly reduce the scan time by avoiding redundant sampling of the same k-space locations across different \gls{ti} blocks.

Although initial \textit{in vivo} results in healthy volunteers demonstrate evident improvements in image sharpness and contrast stability, future work must incorporate formal radiological evaluations and larger cohorts of patients to confirm diagnostic equivalence.
A larger study population is also required to provide us with comprehensive statistical validation of volumetric consistency and morphometry, as demonstrated in~\cite{Hanford2026}, ensuring that increased acceleration does not compromise clinically relevant structural biomarkers.
This quantitative validation will determine whether the nearly-a-minute acquisition paradigm maintains sufficient diagnostic quality in various pathologies and demographics.
However, for a smooth deployment across large cohorts, online reconstruction is crucial to allow radiologists to assess image quality in real time.
Although our reconstructions are surely doable in real time, their integration directly into clinical workflows remains a vital step for immediate diagnostic evaluation.

\section*{Acknowledgements}
The authors thank Gian Franco Piredda for providing to our institution under the Research Software Package the prototypal CS-based TurboFLASH MR sequence with \gls{pds} and \gls{cp} samplings.
The authors thank Franck Mauconduit for helpful discussions regarding the pulse sequence and arbitrary gradients being played on the MR system.
The authors also thank Pierre-Antoine Comby for significant and helpful contributions to the \texttt{mri-nufft} library.
Finally, the authors thank the researchers involved in the scan of the SENIOR cohort~\cite{Haeger2020} at NeuroSpin, during which our acquisitions were optimized.

\FloatBarrier
\bibliography{ref}

\end{document}